\documentclass{article}
\usepackage{spconf,booktabs,amsmath,epsfig}
\usepackage{amssymb}
\usepackage{gensymb}
\usepackage{multirow}
\usepackage{color,array,dcolumn}
\usepackage{fixltx2e}
\usepackage{stfloats}
\usepackage{url}
\usepackage{amssymb}
\usepackage{gensymb} 
\usepackage{multirow} 

\usepackage{color}

\begin{document}

\title{Recurrent Neural Networks with Stochastic Layers for Acoustic Novelty Detection}

\name{Duong Nguyen $^1$ 
\thanks{ This  work  was  supported  by  the UBL Mobility Fund and the Natural Sciences and Engineering Research Council of Canada (NSERC). The authors would like to thank A3Lab for the dataset.}
, Oliver S. Kirsebom $^2$ 
\thanks{Copyright 2019 IEEE. Published in the IEEE 2019 International Conference on Acoustics, Speech, and Signal Processing (ICASSP 2019), 
 https://2019.ieeeicassp.org. 
 }
, F\'abio Fraz\~ao $^2$, Ronan Fablet $^1$, Stan Matwin $^{2,3}$}
\address{$(1)$ IMT Atlantique, Lab-STICC, UBL, Brest, France\\
(2) Institute for Big Data Analytics, Dalhousie University, Halifax, Canada \\
(3) Institute of Computer Science, Polish Academy of Sciences, Warsaw, Poland}

\maketitle

\begin{abstract}

In  this paper, we adapt Recurrent Neural Networks with Stochastic Layers, which are the state-of-the-art for generating text, music and speech, to the problem of acoustic novelty detection. By integrating uncertainty into the hidden states, this type of network is able to learn the distribution of complex sequences. Because the learned distribution can be calculated explicitly in terms of probability, we can evaluate how likely an observation is then detect low-probability events as novel.
The model is robust, highly unsupervised, end-to-end and requires minimum preprocessing, feature engineering or hyperparameter tuning. An experiment on a benchmark dataset shows that our model outperforms the state-of-the-art acoustic novelty detectors. 
%
%

\end{abstract}

\section{Introduction}
\label{secIntroduction}
 Audio processing in general, and acoustic novelty detection in particular has attracted significant attention recently. A number of studies have used acoustic data to detect abnormal events, mostly for surveillance purposes, such as human fall detection \cite{zhuang_acoustic_2009}, \cite{salman_khan_unsupervised_2015}, abnormal jet engine vibration detection \cite{clifton_novelty_2015}, hazardous events detection \cite{ntalampiras_probabilistic_2011}.
 
The main challenge of novelty detection is we do not have a large amount of novel events to learn their characteristics, while the normal set is usually very big and contains a large amount of uncertainty. 
The common approach is to use unsupervised methods to learn the normality model, then consider events that do not fit this model as abnormal (novel). Most of these systems use Gaussian Mixture Model (GMM) or Hidden Markov Model (HMM) \cite{kumar_multimodal_2005}, \cite{ntalampiras_probabilistic_2011}, \cite{atrey_audio_2006}. Bayesian Networks have also been explored \cite{zajdel_cassandra:_2007}, \cite{giannakopoulos_audio-visual_2010}. Recently, advances in deep learning \cite{lecun_deep_2015}, especially in Recurrent Neural Networks (RNNs) and their extensions (Long Short-Term Memory --- LSTM \cite{sak_long_2014}, Gated Recurrent Unit --- GRU \cite{chung_gated_2015}) have opened new venues for acoustic modeling. 
In \cite{marchi_novel_2015}, the authors employed LSTMs to create an AutoEncoder (AE) to model normal sounds and detect abnormal sounds using the reconstruction errors. This idea has been extended in \cite{principi_acoustic_2017} by applying an adversarial training protocol.

However, acoustic signals are stochastic. RNN-based networks, whose hidden states are deterministic, can hardly capture all the variations in the data.
Recent efforts to improve the modeling capacity of RNNs by including stochastic factors in their hidden states have shown impressive 
results, especially for generating text, music and speech \cite{bayer_learning_2014}, \cite{boulanger-lewandowski_modeling_2012}, \cite{chung_recurrent_2015}, \cite{fraccaro_sequential_2016}.

In this paper, we adapt these models to create an unsupervised acoustic novelty detector. 
Our approach performs an end-to-end learning of a probabilistic representation of acoustic signals.  
Given this representation, 
we can evaluate how likely an observation and state the detection of novel events as the detection of observations with a low probability.
We argue that this model is robust, highly unsupervised, end-to-end and requires minimum preprocessing, feature engineering or hyperparameter tuning. Our empirical evaluation on a dataset for novel event detection in audio data shows that the proposed model outperforms the state-of-the-art. 

The paper is organized as follows: in Section \ref{secArchitecture}, we present the details of the proposed approach; we compare the model with state-of-the-art methods to point out its advantages in Section \ref{secRelatedWork}; the experiment and results are shown in Section \ref{secExperimentResult}; finally in Section \ref{secConclusionsPerspectives} we give  conclusions and some perspectives for future work.

\section{The proposed approach}
\label{secArchitecture}

\subsection{Recurrent Neural Networks with Stochastic Layers (RNNSLs)}
\label{secRNNSLs}

For time series modeling, the two most common approaches are State Space Models (SSMs) and Recurrent Neural Networks (RNNs). SSMs such as Kalman filters \cite{brown_introduction_nodate} and particle filters \cite{doucet_tutorial_nodate} have been explored for a long time and are the state-of-the-art model-driven schemes thanks to their ability to model stochasticity. However, these models are limited  by their mathematical assumptions (for example, Kalman filters assume the data generating process is Gaussian). RNNs, on the other hand, have attracted a lot of attentions recently by their capacity to represent long-term dependencies in time series \cite{lecun_deep_2015}. The main drawback of RNNs is that their hidden states are deterministic, making them unable to capture all the stochastic components of the data.
A number of efforts have been made to bring together the power of SSMs and RNNs
\cite{bayer_learning_2014}, \cite{chung_recurrent_2015}, \cite{fraccaro_sequential_2016}, \cite{krishnan_deep_2017}: Recurrent Neural Networks with Stochastic Layers (RNNSLs).

RNNSLs aim to learn the distribution $p$, which can be factored through time, over a sequence of $T$ observed random variables $\{ \boldsymbol{\mathrm{x}}_{t} \} _{, t = 1..T}$:
\begin{equation}
p(\boldsymbol{\mathrm{x}}_{1:T}) = \prod_{t=1}^T p_t(\boldsymbol{\mathrm{x}}_t|\boldsymbol{\mathrm{x}}_{<t}),
\end{equation}
where $\boldsymbol{\mathrm{x}}_{<t}$ denotes $\boldsymbol{\mathrm{x}}_{1:t-1}$.

Following a SSM formulation, we assume that the data generation process of $\boldsymbol{\mathrm{x}}_{1:T}$ relies on a sequence of $T$ latent random variables $\{ \boldsymbol{\mathrm{z}}_{t} \} _{, t = 1..T}$.
At each time step $t$, the joint distribution $p_t(\boldsymbol{\mathrm{x}}_t,\boldsymbol{\mathrm{z}}_t|\boldsymbol{\mathrm{x}}_{<t}\boldsymbol{\mathrm{z}}_{<t})$ can be factored into:
\begin{multline}
p_t(\boldsymbol{\mathrm{x}}_t,\boldsymbol{\mathrm{z}}_t|\boldsymbol{\mathrm{x}}_{<t}\boldsymbol{\mathrm{z}}_{<t}) = p_t(\boldsymbol{\mathrm{x}}_t|\boldsymbol{\mathrm{x}}_{<t},\boldsymbol{\mathrm{z}}_{\leq t})p_t(\boldsymbol{\mathrm{z}}_t|\boldsymbol{\mathrm{x}}_{<t},\boldsymbol{\mathrm{z}}_{<t}), 
\end{multline}
where $\boldsymbol{\mathrm{z}}_{\leq t}$ denotes $\boldsymbol{\mathrm{z}}_{1:t}$. In other words, each time step of the network is an autoencoder, conditionally to the historical information.

Depending on the stochastic nature of the considered data, the emission distribution $p_t(\boldsymbol{\mathrm{x}}_t|\boldsymbol{\mathrm{x}}_{<t},\boldsymbol{\mathrm{z}}_{\leq t})$ may be highly nonlinear. However, this nonlinearity usually leads to the intractability of the inference distribution $p_t(\boldsymbol{\mathrm{z}}_t|\boldsymbol{\mathrm{x}}_{ \leq t},\boldsymbol{\mathrm{z}}_{<t})$. The most common solution to overcome this obstacle is the variational approach \cite{boulanger-lewandowski_modeling_2012}, \cite{chung_recurrent_2015}, \cite{fraccaro_sequential_2016}, which introduces an approximation $q_t(\boldsymbol{\mathrm{z}}_t|\boldsymbol{\mathrm{x}}_{ \leq t},\boldsymbol{\mathrm{z}}_{<t})$ of the posterior distribution $p_t(\boldsymbol{\mathrm{z}}_t|\boldsymbol{\mathrm{x}}_{ \leq t},\boldsymbol{\mathrm{z}}_{<t})$ then estimates $p_t(\boldsymbol{\mathrm{x}}_t|\boldsymbol{\mathrm{x}}_{<t})$  by the Evidence Lower BOund (ELBO) $\mathcal{L}(\mathrm{x},p_t,q_t)$:
\vspace{-0.2cm}
\begin{multline}
    \log p_t(\boldsymbol{\mathrm{x}}_t|\boldsymbol{\mathrm{x}}_{<t}) \geq \mathcal{L}(\mathrm{x},p_t,q_t) = \\
\mathbb{E}_{\boldsymbol{\mathrm{z}}_{t}\sim q_t}\big[\log p_t(\boldsymbol{\mathrm{x}}_t|\boldsymbol{\mathrm{x}}_{<t},\boldsymbol{\mathrm{z}}_{\leq t}) \big] \\
    - \mathrm{KL}\big[q_t(\boldsymbol{\mathrm{z}}_t|\boldsymbol{\mathrm{x}}_{ \leq t},\boldsymbol{\mathrm{z}}_{<t})||p_t(\boldsymbol{\mathrm{z}}_t|\boldsymbol{\mathrm{x}}_{<t},\boldsymbol{\mathrm{z}}_{<t}) \big] 
	\label{eqELBO}
\end{multline}


where $\mathrm{KL}\big[q_t||p_t\big] $ is the Kullback-Leibler divergence between two distributions $q_t$ and $p_t$. 

There are several types of RNNSLs, differing in the way that they model the structure of the latent space. The most common types are Variational Recurrent Neural Networks (VRNNs) \cite{chung_recurrent_2015}, Stochastic Recurrent Neural Networks (SRNNs) \cite{fraccaro_sequential_2016} and Deep Kalman Filters (DKFs) \cite{krishnan_deep_2017}.
We experimented most of these types, however, in this paper, for simplicity purposes, we only report the VRNNs, introduced by Chung et al. \cite{chung_recurrent_2015}.

In VRNNs, the historical information $(\boldsymbol{\mathrm{x}}_{<t},\boldsymbol{\mathrm{z}}_{<t})$ is encoded by the dynamics of the hidden states of their RNN (LSTM) $\boldsymbol{\mathrm{h}}_t = h(\boldsymbol{\mathrm{x}}_{t-1},\boldsymbol{\mathrm{z}}_{t-1},\boldsymbol{\mathrm{h}}_{t-1})$. 
More precisely, it involves the parameterization of the following distributions, namely 
the emission distribution  $p_t(\boldsymbol{\mathrm{x}}_t|\boldsymbol{\mathrm{x}}_{<t},\boldsymbol{\mathrm{z}}_{ \leq t}) = p(\boldsymbol{\mathrm{x}}_t|\boldsymbol{\mathrm{z}}_t, \boldsymbol{\mathrm{h}}_t)$, the prior distribution  $p_t(\boldsymbol{\mathrm{z}}_t|\boldsymbol{\mathrm{x}}_{<t},\boldsymbol{\mathrm{z}}_{<t}) = p(\boldsymbol{\mathrm{z}}_t|\boldsymbol{\mathrm{h}}_t)$ and the variational posterior distribution $q_t(\boldsymbol{\mathrm{z}}_t|\boldsymbol{\mathrm{x}}_{ \leq t},\boldsymbol{\mathrm{z}}_{<t}) = p(\boldsymbol{\mathrm{z}}_t|\boldsymbol{\mathrm{x}}_t,\boldsymbol{\mathrm{h}}_t)$ as neural networks.
Here, we consider fully connected networks with Gaussian formulation of these three distributions. For more details of VRNNs, please refer to \cite{chung_recurrent_2015}.
%
%

\subsection{RNNSLs for Acoustic Novelty Detection}
\label{secRNNSLsforAcousticNoveltyDetection}
RNNSLs were initially designed for generating text, music, speech. They are currently the state-of-the-art in these domains \cite{boulanger-lewandowski_modeling_2012}, \cite{chung_recurrent_2015}, \cite{fraccaro_sequential_2016}, \cite{maddison_filtering_2017}. The interesting point of this type of models in comparison to other state-of-the-art methods like Wavenet \cite{oord_wavenet:_2016} is that these models calculate the distribution $p(\boldsymbol{\mathrm{x}}_{1:T})$ explicitly, so that after learning this distribution from the training set, we can evaluate the probability for each new sequence. The idea of using RNNSLs for novelty detection was first introduced in \cite{nguyen_multi-task_2018} for the detection of abnormal behaviors of vessels, we adapt this model to novelty detection in acoustic data.

\begin{figure}
  \centering
  \includegraphics[width=85mm]{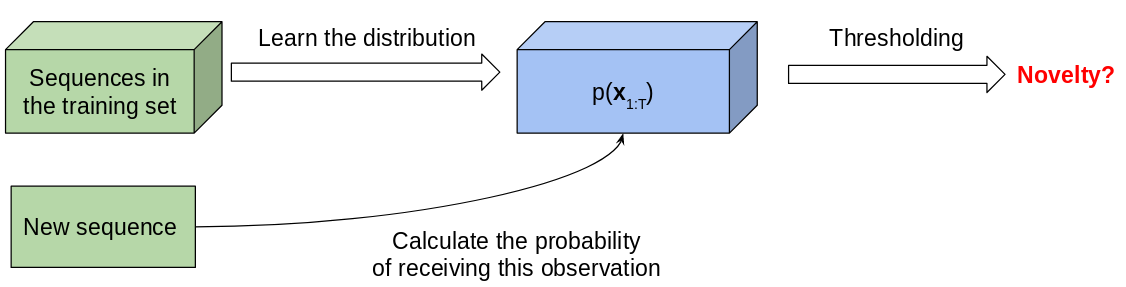}%
  \caption{Architecture of the proposed RNNSL-based novelty detector.} \label{figArchitecture}
\end{figure}

Here, an acoustic signal is modeled as a time series $\{ \boldsymbol{\mathrm{x}}_{t} \}_{, t = 1..T}$  where $\boldsymbol{\mathrm{x}}_t$ can be a chunk of $n$ samples of the waveform, or $n$ frequency bins  in a spectrogram at a given time $t$. A RNNSL first learns the distribution over $\boldsymbol{\mathrm{x}}_{1:T}$ in the training set, which may or may not contain some abnormal sequences. Then, for any new acoustic signal, we can evaluate its log-probability.
 If this log-probability is smaller than a threshold, the sequence will be considered as abnormal (or novel), as illustrated in Fig. \ref{figArchitecture}.
 
To choose the threshold, we create a validation set, which again may or may not contain some abnormal sequences and compute the mean $\mu_{valid}$ and the standard deviation $\sigma_{valid}$ of the log-probability of the sequences in this set. The value of the threshold is then chosen as: $\theta = \mu_{valid} - \alpha*\sigma_{valid}$.
$\alpha$ is usually chosen as $3$.  

The training set and the validation set may contain some abnormal sequences. However, since RNNSLs are probabilistic models, they will eventually ignore these ``outliers" (this conjecture is confirmed experimentally). This property helps to reduce data cleaning efforts.

\begin{figure}
  \centering
  \includegraphics[width=65mm]{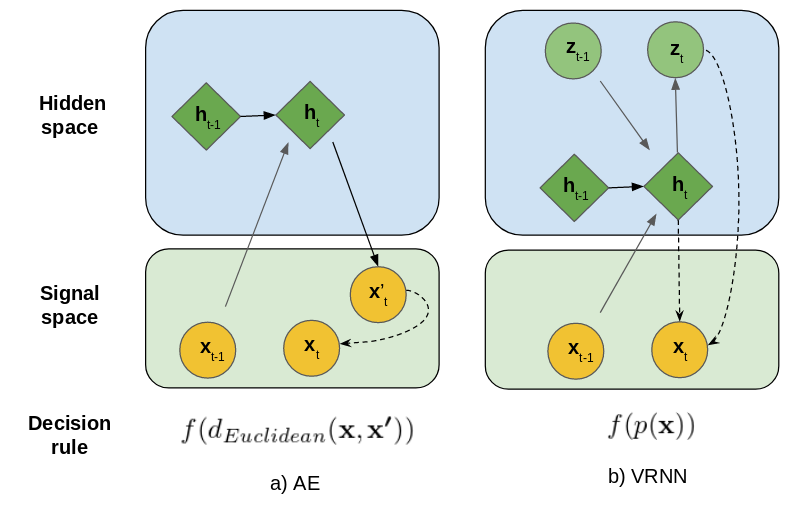}%
  \caption{Architecture and decision rule of the proposed model (VRNN) in compared to previously proposed AE-based models. $\boldsymbol{\mathrm{x}}_t$ is the original signal at the given time step $t$, $\boldsymbol{\mathrm{h}}_t$ is the hidden state of the RNN (LTSM), $\boldsymbol{\mathrm{z}}_t$ is the latent stochastic state, $\boldsymbol{\mathrm{x}}'_t$ is the reconstructed output of the AE. The solid arrows denote the calculation processes, while the dashed arrows show how the cost function is calculated. We use the same notation as \cite{fraccaro_sequential_2016}, circles for stochastic factors, diamonds for deterministic factors.} \label{figRule}
  \vspace{-4mm}
\end{figure}
\section{Related work}
 \label{secRelatedWork}
 
A number of researches have explored deep neural networks to detect novelty in acoustic surveillance. We point out here the advantages of our model over those used in \cite{marchi_novel_2015} and \cite{principi_acoustic_2017}, which are currently the state-of-the-art methods. 

Both \cite{marchi_novel_2015} and \cite{principi_acoustic_2017} used RNNs (LSTMs in particular) as an AutoEncoder (AE) which can reconstruct the original signal from a compressed representation (Compression AutoEncoders --- CAEs) or from a corrupted version of it (Denoising AutoEncoders --- DAEs). However, as discussed in \cite{chung_recurrent_2015}, \cite{fraccaro_sequential_2016} and \cite{krishnan_deep_2017}, the fact that the hidden states of RNNs are deterministic reduces their capacity to capture all data variabilities, especially for data that contain high levels of randomness. 

Moreover, the detection criterion used in \cite{marchi_novel_2015} is the Euclidean distance between the original input and the reconstructed output of the autoencoder. This criterion is very sensitive to noise. 
\cite{principi_acoustic_2017} addressed this drawback by using an adversarial strategy, however, the ultimate idea is also to compare the original input and the reconstructed output from the autoencoder. By contrast, our method detects novel events by directly evaluating the probability of the received signal. Besides the improved detection criterion, the architecture of our model is also more robust to noise \cite{nguyen_multi-task_2018}.  

These differences are sketched in Fig. \ref{figRule}. The hidden space of our model has stochastic factors, which help to increase modeling capacity. The decision rule of our model is a function of the distribution learned by the network, making the model more robust to noise. 

The selection of the thresholding value for novelty detection is another important difference compared to previous works. 
The approach in \cite{marchi_novel_2015} is not fully unsupervised, because it needs some information about the proportion of abnormal events in the data. 
Our method, in contrast, only uses the information from the training set and the validation set to chose the threshold, without any prior knowledge of the annotations, based on a statistically-sound criterion, {\em i.e.} the false alarm rate. 


\section{Experiment and Result}
\label{secExperimentResult}

\subsection{Dataset}
\label{secDataset}

We tested our model\footnote{The code is available at https://github.com/dnguyengithub/AudioNovelty} on the same dataset used in \cite{marchi_novel_2015} and \cite{principi_acoustic_2017}, which is part of the PASCAL CHiME speech separation and recognition challenge dataset \cite{barker_pascal_2013}. The original dataset contains 7 hours of in-home environment recordings with two children and two adults performing common activities, such as talking, eating, playing and watching television. The author of \cite{marchi_novel_2015} took a part of those recordings and created a dataset for acoustic novelty detection (100 minutes for the training set and 70 minutes for the test set). In the new dataset, the sounds of the PASCAL CHiME are considered as background, the test set was generated by digitally adding abnormal sounds like alarms, falls, fractures (breakages of objects), screams. The details of the dataset were presented in \cite{marchi_novel_2015}.

\subsection{Experimental Setup}
\label{secExperimentalSetup}

In order to use the models in \cite{marchi_novel_2015} and \cite{principi_acoustic_2017} as baselines, we set up our model to have the same evaluation metric that was used in those papers. However, instead of transforming the data to mel spectrograms like in \cite{marchi_novel_2015} and \cite{principi_acoustic_2017}, we worked directly with the waveform (end-to-end model). The dataset was recorded by a binaural microphone at a sample rate of 16kHz. We converted each audio to 1 channel and then split it into sequences of 160-dimensional frames, each frame corresponds to 0.01s, as in \cite{marchi_novel_2015} and \cite{principi_acoustic_2017}. \cite{marchi_novel_2015} and \cite{principi_acoustic_2017} evaluated the detection at each frame instead of at the whole sequence, so we also applied the thresholding step to each $\log p(\boldsymbol{\mathrm{x}}_t|\boldsymbol{\mathrm{x}}_{<t})$, instead of $\log p(\boldsymbol{\mathrm{x}}_{1:T})$. 

We tested different topologies of VRNN, with the latent size of 64, 80, 160 and 200.
The models were trained using Adam optimizer \cite{kingma_adam:_2015}, with a learning rate of $3e-5$.

\subsection{Results}
\label{secResult}

Different configurations gave different log-likelihoods on the dataset, however the final detection results were quite similar. We report here only one of the topologies, which gave the best result: VRNN with 160 latent units (the models with 80 hidden units also gave similar results).
We compare the performance of our model with the result of GMM, HMM, those in \cite{marchi_novel_2015} (LSTM-based CAE, LSTM-based DAE) and in \cite{principi_acoustic_2017} (Adversarial AE). The result is shown in  Table \ref{tabF1result}\footnote{The values in Table \ref{tabF1result} are from \cite{marchi_novel_2015} and \cite{principi_acoustic_2017}, \cite{principi_acoustic_2017} did not show the precision and recall of their model}. Besides choosing the threshold automatically as discussed in Section \ref{secArchitecture}, we also used the same technique as in \cite{marchi_novel_2015} to chose the optimal threshold value, denoted as $\textbf{VRNN* }$.

\begin{table}
  \renewcommand{\arraystretch}{1.3}
  \caption[Detection result]{Detection result, in comparison with state-of-the-art methods.
  }\label{tabF1result}
  \vspace{1mm}
  \centering{\footnotesize
  \begin{tabular}{| c | c | c | c | c |}
    \hline
	 \textbf{Method }	&  \begin{tabular}{@{}c@{}} \textbf{Online} \\ \textbf{Processing} \end{tabular} 	& \textbf{Precision} & \textbf{Recall} &\textbf{ F1 score} \\	 
    \hline
	GMM		&  Yes	& 99.1 & 87.8 & 89.4			\\
    \hline
    HMM		&  Yes	& 94.1 & 88.9 & 91.1		\\
	\hline
    LSTM-CAE &  Yes	& 91.7 & 86.6 & 89.1			\\
    BLSTM-CAE &  No	& 93.6 & 89.2 & 91.3			\\
    \hline
    LSTM-DAE &  Yes	& 94.2 & 90.6 & 92.4			\\
    BLSTM-DAE &  No	& 94.7 & 92.0 & 93.4			\\
    \hline
    Adversarial AE & ?	& ?	& ?	& 93.3		\\
    \hline
    \textbf{VRNN }&  Yes & 95.4 & 91.8 &  \textbf{93.6}			\\
    \textbf{VRNN* }&  Yes & 95.4 & 92.8 & \textbf{94.1}			\\
    \hline
  \end{tabular}}
\end{table}

\begin{table}
  \renewcommand{\arraystretch}{1.3}
  \caption{Robustness test.
  }\label{tabRobustness}
  \vspace{1mm}
  \centering{\footnotesize
  \begin{tabular}{| c | c | c | c |}
    \hline
	 \textbf{SNR }	& \textbf{Precision} & \textbf{Recall} &\textbf{ F1 score} \\	 
    \hline
    5dB &   96.0 & 91.2 & 93.6			\\
    \hline
    10dB &   96.1 & 91.9 & 94.0			\\
    \hline
    15dB &   96.1 & 92.1 & 94.0			\\
    \hline
  \end{tabular}}
\end{table}

Our method not only outperformed the state-of-the-art methods, but also has the ability to work online, which is highly beneficial for real-time surveillance. Models that use bidirectional LSTM (BLSTM-CAEs, BLSTM-DAEs) can not reach online processing the because a look-ahead buffer is required. The online processing ability of Adversarial AEs depends on the structure that they use (LSTM or BLSTM). 

\begin{figure}[h!]
  \centering
  \includegraphics[width=83mm]{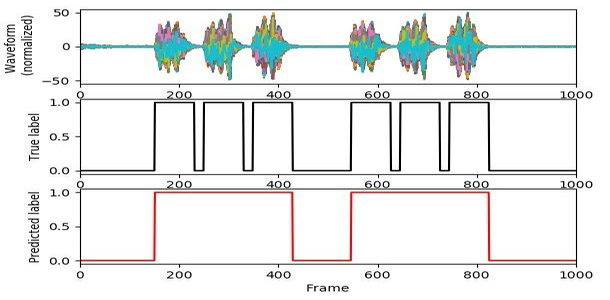}%
  \caption{An example where the novelty events were merged. This figure shows the waveform of two alarms, each alarm consists of there ``beeps'', our model considered this ``beep beep beep" as one event, while the annotation made by the authors of \cite{marchi_novel_2015} separates these ``beeps". } \label{figMerged}
\end{figure}

\begin{figure}[!]
  \centering
  \includegraphics[width=83mm]{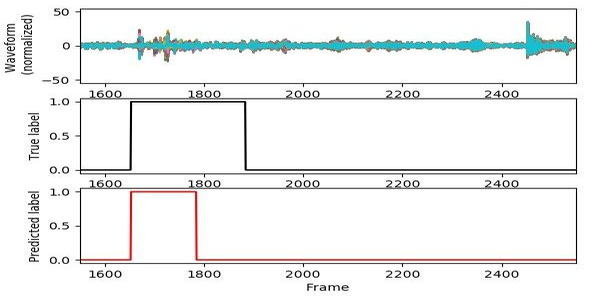}%
  \caption{An example where the model missed a part of the novelty event. This figure shows the waveform of the sound of a fracture of a dish. The tail of the sound is very mall and gradually becomes submerged in the background.} \label{figMiss}
  \vspace{-3mm}
\end{figure}

When investigating the cases where the proposed model misdetected the novelty, we found that actually the model could detect all the novel events, however, the way the detection was evaluated reduced the accuracy. As in \cite{marchi_novel_2015} and \cite{principi_acoustic_2017}, the detection was evaluated at each time step of 0.01s. Our model has a memory effect (the memory of its LSTM cells), so it tends to merge the abnormal events that are very close to each other, as shown in Fig. \ref{figMerged}. In other cases, the model missed a part of the sound, especially for the tail of the fractures, as shown in Fig. \ref{figMiss}. These sounds have a long tail which is gradually submerged in the background. 
These misdetections are not detrimental in real life applications, because we are more interested in whether or not there is a novel event than on how long the event is.


We also conducted a robustness test where we added Gaussian noise to the test set. The additive noise is unknown by the model. This is a common scenario in audio surveillance, when the background environment changes ({\em e.g.} because of winds) or when noise appears in the electronic system. Table \ref{tabRobustness}\footnote{\cite{marchi_novel_2015} and \cite{principi_acoustic_2017} did not provide sufficient detail to replicate their codes for this test.} shows the performance of the proposed approach (with optimal threshold) on the corrupted test sets with different level of Signal to Noise Ratio (SNR). Thanks to the nature of VRNNs and the improved detection criterion, our model is robust to noise.

\section{Conclusions and perspectives}
\label{secConclusionsPerspectives}

We have presented a novel unsupervised end-to-end approach for acoustic novelty detection. This approach exploits RNNs with stochastic layers, which are the state-of-the-art frameworks for time series modeling.
Given the learned probabilistic representations, novelty detection can be stated as a classic statistical test, which fully accounts for the stochasticity of the considered acoustic datasets. Reported experiments on a benchmarked dataset showed that the model outperforms the state-of-the-art detectors \cite{marchi_novel_2015}, \cite{principi_acoustic_2017}.


The dataset used in this paper is quite simple, the novel events in it are quite easy to be detected. Future work could involve applying this model to more complex signals, {\em e.g.} underwater acoustic signals which depict even greater variabilities. The impact of the threshold is also being studied to obtain better threshold selection rule. 



\bibliographystyle{IEEEtran}
\small
\bibliography{Zotero}
\end{document}